\documentstyle[12pt]{article} 
\begin{document}
\bibliographystyle{unsrt}

\def\onehalf{{\textstyle \frac12}}
\def\pr{^\prime}
\def\tsty#1#2{{\textstyle\frac{#1}{#2}}}
\def\ssr#1{{\scriptscriptstyle\rm #1}}
\def\ps{{\partial_\xi}}


\vbox {\vspace{6mm}} 

\begin{center}
{\LARGE \bf MEIXNER OSCILLATORS}\\[7mm]
Natig M.\ Atakishiyev,\footnote{Instituto de Matem\'aticas, UNAM, Apartado
Postal 273-3, 62210 Cuernavaca, Morelos, M\'exico}
Elchin I.\ Jafarov,$^2$ \\
Shakir M.\ Nagiyev,\footnote{Institute of Physics, Azerbaijan Academy
of Sciences, H.\ Javid Prospekt 33, Baku 370143, Azerbaijan, e-mail: azhep@lan.ab.az}
and Kurt Bernardo Wolf \\
{\it Instituto de Investigaciones en Matem\'aticas Aplicadas y en
Sistemas ---Cuernavaca \\ 
Universidad Nacional Aut\'onoma de M\'exico  \\
Apartado Postal 48--3 \\ 62251 Cuernavaca, Morelos, M\'exico}\\[5mm]

\end{center}


\def\s{\sum_{n=0}^\infty}
\def\ss{\sum_{\xi=0}^\infty}
\def\ssp{\sum_{\xi'=0}^\infty}
\def\sk{\sum_{k=0}^\infty}

\def\onehalf{{\textstyle\frac12}}
\newcommand{\be}{\begin{equation}}
\newcommand{\ee}{\end{equation}}
\newcommand{\bea}{\begin{eqnarray}}
\newcommand{\eea}{\end{eqnarray}}

\vspace{2mm}

\begin{abstract}

	Meixner oscillators have a ground state and an `energy' spectrum
that is equal\-ly spaced; they are a two-parameter family of models
that satisfy a Hamiltonian equation with a {\it difference\/}
operator. Meixner oscillators include as limits and particular cases
the Charlier, Kravchuk and Hermite (common quantum-mechanical)
harmonic oscillators.  By the Sommerfeld-Watson transformation they
are also related with a relativistic model of the linear harmonic
oscillator, built in terms of the Meixner-Pollaczek polynomials, and
their continuous weight function.  We construct explicitly the
corresponding coherent states with the dynamical symmetry group
Sp(2,$\Re$). The reproducing kernel for the wavefunctions of these
models is also found. 

\end{abstract}



\section{Introduction}
	An oscillator is called {\it harmonic\/} when its
oscillation period is independent of its energy. In quantum theory,
this statement leads to its characterization by a Hamiltonian
operator whose energy spectrum is discrete, lower-bound, and equally
spaced \cite{Moshinsky},
\be
	H\psi_n=E_n\psi_n,\qquad E_n \sim n+\hbox{constant},\quad
				n=0,1,2,\ldots\ . \label{energy}
\ee
Within the framework of Lie theory, this further indicates that only
a few choices of operators and Hilbert spaces are available if the
Hamiltonian operator is incorporated into some Lie algebra of low
dimension. 

	If we relax the strict Schr\"odinger quantization rule, we find a
family of harmonic oscillator models characterized by Hamiltonians
that are {\it difference\/} operators (rather than {\it
differential\/} operators).  Their spectrum is also (\ref{energy}),
with $n$ either unbounded, or with an upper bound $N$.  The
wavefunctions are still continuously defined on intervals either
unbounded or bounded, but the governing equation will relate their
values only at discrete, equidistant points of space; the Hilbert
space of wavefunctions will then also have discrete measure. Thus
`space' appears to be {\it discrete\/} rather than continuous. The
two-parameter family of Meixner oscillator models, to be examined
here, is harmonic. Limit and special cases of Meixner oscillators
will be shown to include the Charlier, Kravchuk, and
the ordinary Hermite quantum harmonic oscillator models. 

   The Hermite, Charlier and Kravchuk oscillator models are reviewed
in Section 2, together with their limit relations.  Their common
salient feature is to possess associated raising and lowering
operators for the energy quantum number $n$ in (\ref{energy}). In the
first two, Hermite and Charlier, the Hamiltonian operator further
factorizes into the product of these raising and lowering operators;
the relevant Lie algebra is the Heisenberg-Weyl one, which can
be extended to the two-dimensional dynamical symplectic algebra ${\rm
sp}(2,\Re)={\rm so}(2,1)={\rm su}(1,1)$. 

	The Meixner oscillator model is introduced in Section 3, using
well-known properties of the Meixner polynomials and its difference
equation. The dynamical symmetry is also ${\rm sp}(2,\Re)$. It is
then natural to build the coherent states of Meixner oscillators in
Section 4. Section 5 establishes the analogue for Meixner
wavefunctions of the well-known property of the Hermite functions to
self-reproduce under Fourier (and fractional Fourier) transforms.
This property is of great interest in the processing of signals by
optical means \cite{Lohmann}.  Section 6 shows that limits and
special cases of the Meixner oscillator are the Hermite, Charlier,
and Kravchuk oscillators. In this Section we discuss also how the
Meixner oscillator is related the radial part of the nonrelativistic
Coulomb system in quantum mechanics and a relativistic model of the
linear harmonic oscillator, built in terms of the continuous
Meixner-Pollaczek polynomials.


\section{Hermite, Charlier and Kravchuk oscillators}
	In this Section we collect for the reader the basic facts on the
Hermite, Charlier and Kravchuk oscillators.  We introduce their
Hamiltonian operator and wavefunctions, as well as raising and
lowering operators for each oscillator model. 
Finally, we show the limit relations whereby Charlier and Kravchuk
`discrete' oscillators converge both to the quantum-mechanical
(Hermite) harmonic oscillator. 

\subsection{Hermite (quantum-mechanical) oscillator}
	The linear harmonic oscillator in nonrelativistic quantum
mechanics is governed by the well-known Hamiltonian  
  \be  H^\ssr{H}(\xi) = \frac{\hbar \omega}{2} (\xi^2 - \partial_\xi^2) 
		= \hbar \omega \, [ a^+  a + 1/2 ] \,, \label{1.1}  \ee
where $ \xi = \sqrt{m \omega / \hbar} \, x$ is a dimensionless
coordinate ($m$ is the mass and $\omega$ is the angular frequency of a
classical oscillator); we indicate $ \partial_\xi = {d}/{d \xi}$,
and the creation and annihilation operators are defined as
usual: 
  \be  a^+  = \frac{1}{\sqrt 2} ( \xi - \partial_\xi ) \,, \quad
	a = \frac{1}{\sqrt 2} ( \xi + \partial_\xi ) \,, \qquad
		[a, \, a^+] = 1 . \label{1.2}  \ee

	Eigenfunctions of the Hamiltonian (\ref{1.1}) are expressed in
terms of the {\it Hermite\/} polynomials $H_n(\xi)$,
$n=0,1,2,\ldots$\ .  Their explicit form is
  \be
	H_n(\xi) = (2\xi)^n\,_2F_0(-\onehalf n,\onehalf[1-n];-1/\xi^2)
		= n! \sum_{k=0}^{[n/2]} 
		\frac{(-1)^k (2\xi)^{n-2k}}{k!\,(n-2k)!}, \label{Hermite} \ee
where $[n/2]$ is $\onehalf n$ or $\onehalf(n-1)$ according to whether
$n$ is even or odd. 
Hermite polynomials are orthogonal and of square norm $c_n$ under
integration over $\xi\in\Re$, with measure $\rho^\ssr{H}(\xi)\,d\xi$,
where 
   \be \rho^\ssr{H}(\xi)=e^{-\xi^2},\qquad c_n=\sqrt\pi\,2^n n!.
				\label{Hermitenorm} \ee
Therefore, the normalized wavefunctions 
  \be  \psi_n^\ssr H (\xi) = \sqrt{\rho^\ssr{H}(\xi)/c_n}\, H_n(\xi)
	= \frac{1}{\sqrt{\sqrt{\pi} 2^n n!}} \, 
		H_n (\xi)\, e^{-\xi^2/2}\,, \qquad n = 0, 1, 2, \ldots\ ,
		\label{1.3}  \ee 
are orthonormal and complete in the Hilbert space ${\cal L}^2(\Re)$,
commonly used in quantum mechanics, namely
  \be  \int_{-\infty}^\infty \, d \xi \, \psi_n^\ssr H (\xi) \, 
										 \psi_k^\ssr H (\xi) 
			=\delta _{n,k}\,, \qquad
		\sum_{n=0}^\infty \, \psi_n^\ssr H (\xi) \, \psi_n^\ssr H (\xi') = 
					\delta (\xi - \xi') \,. \label{1.5}  \ee
Their corresponding eigenvalues under (\ref{1.1}) define the energy
spectrum of the harmonic oscillator, and are (\ref{energy}) in the
form 
	\be E_n=\hbar\omega(n+\onehalf). \label{HOspectrum} \ee


\subsection{Charlier oscillator}
	A difference (or discrete) analogue of the linear harmonic
oscillator (\ref{1.1}), can be built on the half-line in
terms of the {\it Charlier\/} polynomials $C_n(x;\mu)$, for any fixed
$\mu > 0$ and $n=0,1,2,\ldots$  \cite{1}. Charlier polynomials are
defined as \cite{2,3}   
  \be  C_n (x; \mu) = {}_2\,F_0 (-n, -x; \mu^{-1} ) 
		= \sum_{k=0}^n \frac{({-n})_k\,({-x})_k}{k!\,\mu^k},
							\label{1.6}  \ee
where $(a)_n=\Gamma(a+n)/\Gamma(a)= a(a+1)\cdots(a+n-1)$ is the
shifted factorial and $\Gamma(z)$ is the Gamma function.

The Hamiltonian for the Charlier oscillator model is a difference
operator \cite{1} 
  \be
 H^\ssr{C} (\xi) = \hbar \omega \bigg[ 2 \mu + \onehalf + {\xi}/{h_1}
	-\sqrt{\mu (\mu + 1 +\xi/h_1)} \, 
					e^{h_1 \partial_\xi} 
		       - \sqrt{\mu (\mu + \xi/h_1)} \, 
					e^{-h_1 \partial_\xi} \bigg],
									\label{1.7}  \ee
where by definition
	\be  e^{\pm y\partial_x}\,f(x)=f(x\pm y) \label{firstdifference}
	\ee
is a shift operator by $y$ with the step $h_1 = 1/
\sqrt{2 \mu}$\,. Eigenfunctions of (\ref{1.7}) have  the
same eigenvalues (\ref{HOspectrum}); they are orthogonal with respect
to the weight function 
  \be \rho^\ssr{C} (x) = \frac{e^{-\mu} \, \mu^x}{\Gamma (x+1)},
			\label{1.8}  \ee 
and have the form
  \be \psi^\ssr{C} (\xi) = (-1)^n \sqrt{ \frac{\mu^n}{n!}\,
			\rho^\ssr{C} (\mu + \xi/h_1)}\,\, C_n (\mu + \xi/h_1;\mu)
											\,.\label{psiC}   \ee 
	 It is clear from the definition (\ref{1.6})  that the Charlier
polynomials are self-dual: 
$ C_n (x; \mu) = C_x (n; \mu)$; therefore the Charlier functions
(\ref{psiC})  satisfy {\it two\/} discrete orthogonality relations
  \be  \sum_{k=0}^\infty \, \psi_m^\ssr{C} (\xi_k) \,
				\psi_n^\ssr{C} (\xi_k) 
			= \delta_{m,n} \,, \qquad 
  \sum_{k=0}^\infty \, \psi_k^\ssr{C} (\xi_m) \,\psi_k^\ssr{C} (\xi_n) =
					\delta_{n,m}, \label{1.10}  \ee
where $\xi_k = (k - \mu)h_1$. These are the discrete analogues of
the continuous orthogonality and the completeness relations in
(\ref{1.5}).

		As in the nonrelativistic case (\ref{1.1}), it is possible to
factorize  \cite{1} the Hamiltonian (\ref{1.7})
  \be H^\ssr{C} (\xi) = \hbar \omega \, ( b^+ \, b + \onehalf ),
									\label{1.11}  \ee
by means of the difference operators
  \be  b  =  \sqrt{\mu + 1 + \xi/h_1} \, e^{h_1 \partial_\xi} - \sqrt\mu , \, 
\qquad   b^+  =  \sqrt{\mu + \xi/ \hbar_1} \, e^{-h_1 \partial_\xi} - 
\sqrt\mu \,. \label{1.12}  \ee
These operators satisfy the Heisenberg commutation relation
   \be [ b , b^+ ] = 1, \label{1.13}  \ee
and their action on the wavefunctions (\ref{psiC}) is 
  \be  b \, \psi^\ssr{C}_n(\xi) = \sqrt n \, \psi_{n-1}^\ssr{C} (\xi) 
		\,, \qquad
   b^+ \, \psi^\ssr{C}_n (\xi) = \sqrt{n+1} \, \psi_{n+1}^\ssr{C} (\xi) \,.
					\label{1.14}  \ee

\subsection{Kravchuk oscillators}
	Another difference analogue of the harmonic oscillator \cite{4} 
can be built on the finite interval $[0, N]$, where $N$ is some
positive integer, in terms of the Kravchuk polynomials \cite{2,3}
  \be K_n (x; p, N) = {}_2F_1 (-n, -x; -N; p^{-1} ) 
			= \sum_{k=0}^n \frac{({-n})_k\,({-x})_k}{k!\,({-N})_k\,p^k\,}.
					\label{1.17}  \ee
This is a family of polynomials, parametrized by $ 0 < p < 1$, of
degree $n=0,1,2,\ldots$ in the variable $x$.

	The corresponding Kravchuk oscillator Hamiltonian is 
a difference operator with step $h_2 = \sqrt{2 N p q} $ \cite{4},
\bea
    H^\ssr{K} (\xi) &=& \hbar \omega \bigg\{ 2 p(1-p) N + \onehalf 
			+ (\onehalf-p)\, \xi/h_2 \nonumber\\
 & &{\qquad}- \sqrt{p(1-p)}\, \left[ \alpha (\xi) \, e^{h_2 \partial_\xi} +
		\alpha (\xi - h_2 ) \, e^{-h_2 \partial_\xi} \right]
		\bigg\}\,, \label{1.18}  \\
	\alpha (\xi) &=& \sqrt{(q N - \xi/h_2 ) ( p N + 1 + \xi/h_2 ) }\,.
		\nonumber \eea
The energy spectrum is the same as in (\ref{HOspectrum}), except that in the
Kravchuk case there are only a finite number of energy levels $ n =
0, 1, ..., N$. 

	The Kravchuk polynomials are orthogonal with respect to the
binomial measure 
  \be  \rho^\ssr{K} (x) = C_n^x \, p^x \,q^{N-x},\qquad
		 C_N^x ={N!}/{ \Gamma (x+1) \, \Gamma(N - x + 1)}\,.
                         \label{1.20}  \ee   
The eigenfunctions of the
difference operator (\ref{1.18}) are   
  \be  \psi_n^\ssr{K} (\xi) = (-1)^n 
			\sqrt{C_N^n \Big(\frac{p}{1-p}\Big)^n
			\, \rho^\ssr{K} (p N + \xi/h_2 )} \, 
				K_n (p N + \xi / h_2; p, N )\,, \label{1.19}  \ee
where $C_n^m$ is the binomial coefficient. 
The Kravchuk polynomials (\ref{1.17}) are also self-dual, and
therefore the Kravchuk functions (\ref{1.19})  satisfy the discrete
orthogonality and completeness relations over the points $\xi_j = ( j
- p N ) h_2 $: 
  \be \sum_{j=0}^N \psi _n^\ssr{K} (\xi_j) \, 
				   \psi_k^\ssr{K} (\xi_j ) = \delta _{n,k}\,,\qquad
      \sum_{j=0}^N \psi_j^\ssr{K} (\xi_n) \, 
				\psi_j^\ssr{K} (\xi_k ) = \delta _{n,k}\,,
								\label{1.22}  \ee
for $n, k = 0, 1, ..., N$. 

	Now, it has been shown in \cite{4} that the difference operators 
  \bea  A (\xi) &=& (1-p)\, \alpha (\xi) \, e^{h_2 \partial_\xi} -
		p e^{-h_2 \partial_\xi}\, \alpha (\xi) +
		\sqrt{p (1-p)} \left[ (2p - 1) N + 2 \xi/h_2\right]\,,
								\label{1.23a} \\  
    A^+ (\xi) &=& (1-p)  e^{-h_2 \partial_\xi} \,\alpha (\xi) -
			p\, \alpha (\xi)\, e^{h_2 \partial_\xi} +
			\sqrt{p(1-p)} \left[ (2p - 1) N + 2 \xi/h_2\right]\,,
								\label{1.23b}  \eea  
together with the operator 
  \be  A_0 (\xi) = \frac{1}{\hbar \omega} 
		\left [ H^\ssr{K} (\xi) - \onehalf(N+1) \right ]\,, \label{1.24}  \ee
close under commutation as the algebra so(3) of the rotation group,
\be
	[A_0,A]=-A,\quad [A_0,A^+]=A^+,\quad [A^+,A]=2A_0.
				\label{so3comrel}
\ee
The action of the operators (\ref{1.23a}--\ref{1.23b}) on the wavefunctions
(\ref{1.19}) is given by 
  \bea  A(\xi) \, \psi_n^\ssr{K}(\xi) &=& 
						\sqrt {n (N - n+ 1)}\,
						\psi_{n-1}^\ssr{K}(\xi)\,,\label{1.25a} \\
       A^+(\xi) \, \psi_n^\ssr{K}(\xi) &=& 
						\sqrt {(n+1) (N - n)}\,
						\psi_{n+1}^\ssr{K}(\xi)\,.
								\label{1.25b}  \eea
We note that the Kravchuk oscillator was applied recently in finite
(multimodal, shallow) waveguide optics \cite{fractionalF-K}.

\subsection{Limiting cases}
	Among the previous models, the Kravchuk oscillator is the most
general; it limits to the 
Charlier oscillator; in turn, the latter limits to the
common Hermite harmonic oscillator \cite{4}:

\noindent{\bf Kravchuk $\longrightarrow$ Charlier}.\quad
Because of the limit relation \cite{3}
  \be \lim_{N \to \infty} K_n (x; \mu/N , N ) 
			= C_n (x; \mu) \label{1.26}  \ee
between the Kravchuk (\ref{1.17}) and Charlier (\ref{1.6})
polynomials, when  $N \to \infty$ and $p = \mu / N \to 0$, the
operators $H^\ssr{K}(\xi)$, $A(\xi)/ \sqrt N $ and $A^+(\xi) / \sqrt
N$ reduce to the Charlier Hamiltonian (\ref{1.11}), and the lowering
and raising operators (\ref{1.12}) for the Charlier functions,
respectively. The  so(3) algebra (\ref{so3comrel}) in turn contracts 
to the Heisenberg-Weyl algebra (\ref{1.13}).

\noindent{\bf Charlier $\longrightarrow$ Hermite}.\quad
In the limit when the Charlier parameter $\mu$ tends to infinity, we
have \cite{3} 
  \be \lim_{\mu \to \infty} \, h_1^{-n} \, C_n (\mu + \xi/h_1; \mu) = 
				(-1)^n \, H_n (\xi) \,. \label{1.15}  \ee
Similarly, in the limit $\mu\to\infty$, the operators (\ref{1.12})
become $b \to a$,  $ b^+  \to a^+$, and $H^\ssr{C}(\xi) \to
H^\ssr{H}(\xi)$. The Charlier functions (\ref{1.8}) coincide then
with the  Hermite functions (\ref{1.3}), {\it i.e.}
  \be \lim_{\mu \to \infty} \, h_1^{-1/2} \, 
			\psi_n^\ssr{C} (\xi) = \psi_n^\ssr{H} (\xi) \,. 
								\label{1.16}  \ee

\noindent{\bf Kravchuk $\longrightarrow$ Hermite}.\quad
From the limit relations \cite{3,4}
  \be  \lim_{N \to \infty} \, (-1)^n \, \sqrt{c_N^n (p/q)^n} \, 
	K_n (p N + \xi/h_2; p, N) = \frac{1}{\sqrt{2^n n!}} \, H_n (\xi)\,, 
						\label{1.27a}  \ee
and
  \be \lim_{N \to \infty} h_2^{-1} \, \rho^\ssr{K} (p N + \xi/h_2) = 
		\frac{1}{\sqrt\pi} \, e^{- \xi^2} \,, \label{1.27b}  \ee
it  follows that 
  \be \lim_{N \to \infty} h_2^{-1/2} \, \psi_n^\ssr{K} (\xi) 
			= \psi_n (\xi)\,. \label{1.28}  \ee
Also, when $N \to \infty$, the operators $H^\ssr{K}(\xi)$, $ A(\xi) /
\sqrt N$ and $A^+(\xi)/ \sqrt N$ reduce to the Hermite Hamiltonian
(\ref{1.1}), annihilation $a(\xi)$, and creation $a^+ (\xi)$ 
operators (\ref{1.2}) for the ordinary quantum harmonic oscillator,
respectively. The so(3) algebra (\ref{so3comrel}) of this finite
oscillator contracts to the Heisenberg-Weyl algebra of quantum
mechanics. 


\section{Meixner oscillators} 
	We now organize the properties of the Meixner polynomials
\cite{2,3} according to the scheme followed in the previous Section.
Known orthogonality relations for the Meixner polynomials lead to
orthonormal functions and a difference Hamiltonian operator, whose
spectrum is the set of energy levels (\ref{energy}).

\subsection{Meixner polynomials and functions}
The Meixner polynomials \cite{2,3} are Gauss hypergeometric
polynomials 
  \be  M_n (\xi; \beta, \gamma) = {}_2F_1 (-n, -\xi; \beta; 1 - 1/\gamma)
		= M_\xi (n; \beta, \gamma)\,. \label{2.1}
  \ee
They form a two-parameter family of polynomials, for $\beta > 0$ and
$ 0 < \gamma < 1$, of degree $n=0,1,2,\ldots$. Their orthogonality
relation is
  \be  \sum_{m=0}^\infty \, \rho^\ssr{M} (m) \, 
		M_n(m; \beta, \gamma) \, M_k(m; \beta, \gamma) = 
			d_n \, \delta_{nk} \label{2.2}  \ee
with respect to the weight function and square norm
  \be  \rho^\ssr{M}(\xi) = \frac{(\beta)_\xi \, \gamma^\xi}{\xi!} \,, \qquad 
		d_n = \frac{n!}{\gamma^n(\beta)_n (1 - \gamma)^\beta}. \label{2.3}  \ee

\def\p{\psi^\ssr{M}}

Hence, the wavefunctions of the form
  \be \psi_n^\ssr{M} (\xi; \beta, \gamma) = 
			(-1)^n \sqrt{\rho^\ssr{M}(\xi)/d_n} \, 
					M_n(\xi; \beta, \gamma) \,, \label{2.6}  \ee
satisfy the discrete orthogonality relations 
  \be \sum_{\xi=0}^\infty  \p_n (\xi; \beta, \gamma) \, 
		\p_k(\xi; \beta, \gamma) = \delta_{n,k}\,,\quad
   \s \p_n(\xi; \beta, \gamma) \, \p_n (\xi'; \beta, \gamma) 
			= \delta_{\xi,\xi'} \,, \label{2.9}  \ee
as a consequence of (\ref{2.2}) and the self-duality of Meixner
polynomials (\ref{2.1}). Henceforth we shall supress for brevity the
super-index {\sc m} from all operators and functions of the Meixner
oscillator model. 

	The Meixner polynomials (\ref{2.1}) satisfy the three-term recurrence
relation \cite{3}
  \be  [n + (n + \beta) \gamma - (1 - \gamma) \xi] M_n (\xi; \beta, \gamma) = 
		(n + \beta) \gamma M_{n+1} (\xi; \beta, \gamma) 
			+ n M_{n-1} (\xi; \beta, \gamma), \label{2.4}  \ee 
and the difference equation in the real argument 
  \be [\gamma (\xi + \beta) e^{\partial_\xi} + 
		\xi e^{-\partial_\xi} - (1 + \gamma)(\xi + \beta/\gamma) +
		(1 + \gamma)(n + \beta/2)] \, M_n(\xi; \beta, \gamma) = 0 \,.
								\label{2.5}  \ee
Hence, the functions (\ref{2.6}) are eigenfunctions of the difference
Meixner Hamiltonian operator 
  \bea H(\xi) & = & \displaystyle
		\frac{1+\gamma}{1-\gamma}\, (\xi + \onehalf\beta) 
			 - \frac{\sqrt{\gamma}}{1 - \gamma} 
	\left [ \mu(\xi) e^{\partial_\xi} + \mu(\xi-1) e^{-\partial_\xi}
	\right] \,,	\label{2.7}			\\
   \mu(\xi) & =& \sqrt{(\xi + 1)(\xi + \beta)} \,, \label{2.7p}  \eea
with eigenvalues 
\be
	E_n= n + \onehalf\beta\,,\qquad n = 0, 1, 2, \ldots\ .
	\label{EnergyM} 
\ee

\subsection{Dynamical symmetry algebra ${\rm Sp}(2,\Re)$}
As in all previous cases, we can construct the dynamical symmetry algebra
(see, for example, \cite{5,6}) by factorizing \cite{7,8} the difference 
Hamiltonian (\ref{2.7}). Indeed, one can verify that
  \be H(\xi) = B \, B^+ + \onehalf\beta - 1 , \label{2.10}  \ee
where $B=B(\xi)$ and $B^+=B^+(\xi)$ are the difference operators
  \bea
	B &=& \displaystyle  \frac{1}{\sqrt{1 - \gamma}} \, 
			\left(\sqrt{\xi+1}\, e^{\frac{1}{2}\ps} -
		\sqrt{\gamma(\xi + \beta - 1)} e^{-\frac{1}{2} \ps}\right) \,, 
						\label{2.11a}  \\
	B^+ &=& \displaystyle  \frac{1}{\sqrt{1 - \gamma}} \, 
			\left( e^{-\frac{1}{2}\ps} \,\sqrt{\xi+1} -
		e^{\frac{1}{2} \ps} \,\sqrt{\gamma(\xi + \beta - 1)} \right)\,. 
						\label{2.11b}  \eea

It is essential to note that the factorization of the Hamiltonian
(\ref{2.7}), in contrast to the case of the harmonic oscillator
(\ref{1.1}) and the difference model (\ref{1.11}), does not lead
immediately to a closed algebra consisting of $H(\xi)$,  $B$ and
$B^+$. To obtain such an algebra, we compute the explicit form of the
commutator between the last two, 
  \bea  [B, B^+] &=&\displaystyle   H(\xi) - \frac{1 + \gamma}{1 - \gamma} \, 
	\Big(\xi + \onehalf\beta \Big) + \onehalf+  \nonumber \\
   {}&+&\displaystyle \frac{\sqrt\gamma}{1 - \gamma} 
		\left( e^{\frac{1}{2} \ps}
	\sqrt{\frac{\xi + \beta - 1}{\xi + \beta}}\, \mu(\xi) \, e^{\frac{1}{2} \ps} +
	e^{-\frac{1}{2} \ps}  \mu(\xi) \, \sqrt{\frac{\xi + \beta - 1}{\xi + \beta}}
	\,  e^{-\frac{1}{2} \ps}\right)\,. \label{2.12}  \eea
The right-hand side of (\ref{2.12}) suggests that it is necessary to
introduce new  operators
  \bea 
	C &=& \displaystyle  \sqrt{\frac{1 - \gamma}{\gamma}} \, 
				B \, e^{- \frac{1}{2} \ps} \sqrt{\xi + 1} 
	= \frac{\xi + 1}{\sqrt{\gamma}} - e^{-\ps} \mu(\xi)  \,, \label{2.13a}\\
  C^+ &=& \displaystyle  \sqrt{\frac{1 - \gamma}{\gamma}} \,\sqrt{\xi+1}\, 
					e^{ \frac{1}{2}\ps} B^+ 
		= \frac{\xi + 1}{\sqrt{\gamma}} - \mu(\xi) e^{\ps} \,. 
					\label{2.13b}  
\eea
These new operators have the following commutation relations with the
Hamiltonian operator
\bea 
	\ [H,C]&=& \displaystyle  -C + \frac{1}{\sqrt{\gamma}} \,
		 ( H + 1 - \onehalf\beta)\,, \label{2.14a}  \\
     \ [H,C^+]&=& \displaystyle  C^+ - \frac{1}{\sqrt{\gamma}} \,
		 ( H + 1 - \onehalf\beta)\,. \label{2.14b}  \eea

	We now build the difference operators
  \bea K_+ &=&\displaystyle   C^+ - \frac{1}{\sqrt \gamma} \,
				( H + 1 - \onehalf\beta ) \nonumber \\
    &=&\displaystyle  \frac{\gamma}{1 - \gamma} \, \mu(\xi) e^{\ps} 
		+ \frac{1}{1 - \gamma}\, e^{- \ps} \mu(\xi) - 
		\frac{2 \sqrt \gamma}{1 - \gamma} (\xi + \onehalf\beta) 
				\,, \label{2.15a} \\ 
	K_- &=&\displaystyle   C - \frac{1}{\sqrt \gamma} \,
			( H + 1 - \onehalf\beta )  \nonumber \\
       &=&\displaystyle  \frac{1}{1 - \gamma} \, \mu(\xi) e^{\ps} 
			+ \frac{\gamma}{1 - \gamma}\, e^{- \ps} \mu(\xi) - 
			\frac{2 \sqrt \gamma}{1 - \gamma} (\xi + \onehalf\beta)
					\,. \label{2.15b}  \eea 
Together with $K_0 = H $, they now form the closed Lie algebra sp(2,$\Re$),
  \be \left[ K_0 , K_{\pm} \right ] = \pm K_{\pm}  \,, \qquad 
		\left[ K_-, K_+ \right ] = 2 K_0 \,. \label{2.16}  \ee
The raising and lowering operators $K_+ $ and $K_-$ are connected by
with the cartesian generators
  \bea
	K_1 =-i\onehalf(K_+-K_-)  &=&\displaystyle  i\onehalf
					[ \mu(\xi) e^{\ps} - e^{-\ps}
					\mu(\xi)] \,,  	\label{2.17-1} \\
	K_2 =-\onehalf(K_++K_-)  &=&\displaystyle  
			- \frac{1 + \gamma}{2 (1 - \gamma)}
				[\mu(\xi) e^{\ps} + e^{-\ps} \mu(\xi)] + \frac{2
				\sqrt \gamma}{1 - \gamma}  
		( \xi + \onehalf\beta )\,. \label{2.17-2}  \eea
The invariant Casimir operator in this case is 
  \be  K^2 = K_0^2 - K_1^2  - K_2^2  = K_0^2 - K_0  - K_+ K_- =
	\onehalf\beta (\onehalf\beta - 1 ) I \,. \label{2.18}  \ee 
The eigenvalue $-\onehalf\beta$ of the Casimir operator $K^2$
determines that the model realizes the unitary irreducible
representation $D^+ (- \beta /2)$ of the Sp(2,$\Re$) group.
The eigenvalues of the compact generator $K_0 (\xi)$ in such
representations are bounded from below and equal to $\onehalf\beta +n
, \, n = 0, 1, 2, \ldots $.  
In other words, a purely algebraic approach enables one to find the correct 
spectrum of the Hamiltonian $H(\xi) = K_0 (\xi)$ in (\ref{EnergyM}).

	The action of the raising and lowering difference operators $K_+  $ and 
$K_-$ on the wavefunctions (\ref{2.6}) is given by
  \be K_+ \, \psi_n(\xi; \beta, \gamma) 
			= \kappa_{n+1} \, \psi_{n+1} (\xi; \beta, \gamma)  \,,\quad
	  K_- \, \psi_n(\xi; \beta, \gamma) 
			= \kappa_{n} \, \psi_{n-1} (\xi; \beta, \gamma)\,, 
										\label{2.19}  \ee
where $\kappa_n = \sqrt{n (n + \beta - 1)}$. Hence the functions $\psi_n
(\xi; \beta, \gamma) $  can be obtained by $n$-fold application of
the operator $K_+$ to the ground  state wavefunction
   \bea
		\psi_n (\xi; \beta, \gamma) 
		&=& \frac{1}{\sqrt{ n! (\beta)_n}} \, K_+^n  \, 
				\psi_0 (\xi; \beta, \gamma)\,, \label{2.20a} \\
		\psi_0 (\xi;\beta, \gamma) 
			&=& \sqrt{(1 - \gamma)^{\beta} \, \rho (\xi)}\,.
					  \label{2.20b}  \eea

\subsection{Unitary equivalence in the second parameter}
	Observe that the eigenvalues of the Casimir operator (\ref{2.18}), as
well as the matrix elements (\ref{2.19}) of the operators $K_+ $ and
$K_- $, do not depend on the second parameter, $\gamma$, of the
Meixner wavefunctions. Therefore the basis functions (\ref{2.6}), 
corresponding to two distinct values of the parameter $\gamma$, must
be intertwined by a unitary transformation. To find its explicit form
we may compare two sets of the generators $K_0 , \, K_1 $
and $K_2 $ [see formulas (\ref{2.7}) and
(\ref{2.17-1},\ref{2.17-2})], corresponding to different values 
$\gamma$ and $\gamma'$. 

	Introducing angles $\theta$ and $\theta'$ such that 
$\gamma = \tanh ^2 \onehalf\theta$, 
$\gamma' = \tanh ^2\onehalf\theta'$  and  $\delta = \theta' - \theta$, 
the relation between the two sets of generators is written  as   
  \bea K_0  &=& \cosh \delta \, K_0' + \sinh \delta \, K_2'  \,,\nonumber\\
		K_1   &=&  K_1'  \,,\hfil \label{2.21}  \\
		K_2   &=&  \sinh \delta \, K_0' + \cosh \delta \, K_2' \,.
						\nonumber  \eea
This shows they are related by a boost in the $0-2$ plane by the
hyperbolic angle  $\delta\in\Re$. Consequently, the wavefunctions
(\ref{2.6}) with different values of the parameter $\gamma$ are
connected by  
  \be 
	\psi_n (\xi; \beta, \gamma) 
		= e^{i K_1' \delta } \, \psi_n(\xi; \beta, \gamma')
		= \sum_{k=0}^\infty \, M^{\gamma,\gamma'}_{n,k} \, 
			\psi_k (\xi; \beta, \gamma') \,.
					\label{2.23}  \ee
The last expression is the matrix form, with elements 
  \bea
	M^{\gamma,\gamma'}_{n,k} &=& \displaystyle \sum_{\xi=0}^\infty \, 
		\psi_n(\xi; \beta, \gamma)\, \psi_k (\xi; \beta, \gamma') \nonumber\\
		&=& \displaystyle (-1)^k \sqrt{\frac{(\beta)_n (\beta)_k}{n!\, k!}} \, 
			(\tanh \onehalf\delta )^{n+k} \, 
			(\cosh \onehalf\delta )^{-\beta} \, 
			M_n (k, \beta, \tanh^2 \onehalf\delta )\,. 
									\label{2.24}  \eea
In deriving (\ref{2.24}) we have used the addition formula for the
Meixner polynomials (\ref{2.1}) given in Ref.\  \cite{9}, Eq.\ (A.6).


\section{Coherent states}
The dynamical symmetry of the Meixner oscillator model (\ref{2.7}),
allows us to construct two kinds of coherent states \cite{10,11}.
Recall that in the case of harmonic oscillator (\ref{1.1}) coherent
states are defined as eigenstates of the annihilation operator $a(\xi)$
\cite{Glauber}. Coherent states for the model (\ref{2.7}) can be
defined either as eigenstates \cite{10} of the  lowering  
operator $K_- (\xi)$, or by acting on the ground state (\ref{2.20b})
with the operator $\exp[\zeta K_+(\xi)]$ \cite{11}. This
gives rise to two distinct coherent states.

\subsection{The Barut-Girardello coherent states}
Characterizing the Barut-Girardello coherent states by the
complex number $z\in{\cal C}$, which is the eigenvalue under the
lowering operator, 
  \be  K_-  \, \phi_z (\xi; \beta, \gamma) 
		= z \, \phi_z (\xi; \beta, \gamma)\,,\label{3.1}  \ee
these coherent states can be expanded in terms of the wavefunctions
(\ref{2.6}),
  \be \phi_z (\xi; \beta, \gamma) 
		= \s \, \frac{z^n}{\sqrt{n! (\beta)_n}} \, 
\psi_n (\xi; \beta, \gamma)\,. \label{3.2}  \ee
Using the generating function \cite{3} for the Meixner polynomials
 (\ref{2.1})
  \be  \s \, \frac{t^n}{n!} \, M_n (\xi; \beta, \gamma) = e^t \,
	{}_1F_1 (-\xi; \beta; \frac{1-\gamma}{\gamma} t),  \label{3.4}  \ee
their explicit form is found to be
  \be \phi_z (\xi; \beta, \gamma) = e^{-z \sqrt \gamma} \, 
		{}_1F_1 \left[-\xi; \beta; \frac{\gamma-1}{\sqrt \gamma} z \right]
			\, \psi_0 (\xi; \beta, \gamma)\,. \label{3.3}  \ee
These coherent states are overcomplete and therefore 
nonorthogonal, 
  \be  \sum_{\xi=0}^\infty 
	\phi_z^*  (\xi; \beta, \gamma)\, \phi_{z'}(\xi; \beta, \gamma) =  
				(z^* z')^{(1-\beta)/2} \, \Gamma (\beta) \, 
				I_{\beta -1} (\sqrt{z^* z'} ) \,,\label{3.5}  \ee
where $I_\nu (z)$ is the modified Bessel function. 

\subsection{Perelomov coherent states}
The second definition of generalized coherent states is due to
Perelomov \cite{11}; it is built through the action of the group
operator $\exp (\zeta K_+)$ on the ground state
$\psi_0 (\xi; \beta,\gamma)$: 
  \bea \chi_\zeta (\xi; \beta, \gamma)
	&=& (1 - |\zeta|^2)^{\beta/2} \, \exp (\zeta K_+) \, 
			\psi_0 (\xi; \beta, \gamma)  \hfil\nonumber \\
	&=&\displaystyle (1 - |\zeta|^2)^{\beta/2} \, \s
			\sqrt{\frac{(\beta)_n}{n!}}  
			\, \zeta^n \, \psi_n (\xi; \beta, \gamma) \,,
			\label{3.6}   \eea
where $\zeta$ is a complex number such that $|\zeta| < 1$. 

	Using the generating function for the Meixner polynomials
\cite{2,3},
  \be \s \frac{(\beta)_n}{n!} \, t^n \, M_n (\xi; \beta, \gamma) 
		= \Big(1 - \frac{t}{\gamma}\Big)^\xi \, 
				(1 - t)^{-\xi-\beta}, \label{3.7}  \ee
we find 
  \be \chi_\zeta (\xi; \beta, \gamma) 
	= (1 - |\zeta|^2)^{\beta/2} \, 
		\Big(1+\frac{\zeta}{\sqrt\gamma} \Big)^\xi \,
		(1 + \sqrt \gamma\, \zeta )^{-\xi - \beta} \, 
		\psi_0 (\xi; \beta, \gamma) \,. \label{3.8}  \ee
These coherent states satisfy the relation [{\it cf.} (\ref{3.5})]
  \be  \sum_{\xi=0}^\infty \chi_\zeta^* (\xi; \beta, \gamma) \, 
			\chi_{\zeta'} (\xi; \beta, \gamma) 
			= [ (1 - |\zeta'|^2) \, (1 - |\zeta|^2)]^{\beta/2} \, 
			(1 - \zeta^* \zeta')^{-\beta} \,.\label{3.9}  \ee


\section{Reproducing transforms}
	Consider the task to find a reproducing kernel for the Meixner functions 
(\ref{2.6}), defined by the relation \cite{21}
  \be   \ssp {\cal K}_t (\xi, \xi') \, \psi_n (\xi', \beta, \gamma) 
	= t^n \, \psi_n (\xi; \beta, \gamma) \,. 
					\label{4.1}  \ee
The quantum mechanical analogue of this expression is the property of
Hermite functions to reproduce under fractional Fourier transforms of
angle $\tau$ for $t=e^{i\tau}$; the common Fourier transform of
kernel $\exp( i\xi\xi')$ corresponds to $\tau=\frac12\pi$
\cite{KBWbook}. The finite-difference Fourier-Kravchuk
transform has the same property on the Kravchuk functions, and has
been shown recently to apply to shallow multimodal waveguides with a
finite number of sensors \cite{fractionalF-K}. 

	 Using the dual orthogonality relation of the Meixner functions
(\ref{2.9}), the explicit form of the kernel ${\cal K}_t (\xi, \xi')$
is found for $|t| < 1$, 
  \be  {\cal K}_t (\xi, \xi') 
	= \s \, t^n \, \psi_n (\xi; \beta, \gamma) \, 
			\psi_n (\xi'; \beta, \gamma) \,.
							\label{4.2}  \ee
It is a bilinear generating function for the Meixner functions.
By the definition (\ref{4.2}), the reproducing kernel ${\cal K}_t (\xi,
\xi') $ is symmetric with respect to exchange of $\xi$ and $\xi'$,
and because of the orthogonality relation (\ref{2.9}) it has the
property 
  \be \ssp {\cal K}_t (\xi, \xi') \, {\cal K}_{t'} (\xi', \xi'') 
			= {\cal K}_{tt'} (\xi, \xi'')\,. 
							\label{4.3}  \ee
Reproducing kernels for the Charlier (\ref{1.8}) and Kravchuk
(\ref{1.19}) functions have been discussed in \cite{12}, whereas the
cases of the $q$-Hermite and Askey--Wilson polynomials have been
considered in \cite{13} and \cite{14,15}, respectively.

Substituting (\ref{2.6}) in (\ref{4.2}), we can write 
  \be  {\cal K}_t (\xi, \xi') = \sqrt{\rho (\xi)\, \rho (\xi')} \, 
		(1 - \gamma)^\beta \,
		\s \, \frac{(\beta)_n}{n!} \, ( \gamma t)^n \, 
		M_n (\xi; \beta, \gamma)\, M_n(\xi'; \beta, \gamma) \,. 
						\label{4.4}  \ee
The sum over $n$ in (\ref{4.4}) is the bilinear generating function
(Poisson kernel) for the Meixner polynomials \cite{16},
  \bea
	&\ & \displaystyle \s \, \frac{(\beta)_n}{n!} \, t^n \, 
  M_n (\xi; \beta, \gamma)\, M_n(\xi'; \beta, \gamma) \nonumber \\
    &\ & \displaystyle \quad{} = (1 - t)^{-\beta-\xi-\xi'} \, 
		( 1 - t/\gamma)^{\xi+\xi'} 
		\, {}_2F_1 \left [ -\xi, -\xi'; \beta; 
			\frac{t (1 - \gamma^2)}{(t - \gamma)^2} \right ].
							\label{4.5}  \eea 
Thus the kernel ${\cal K}_t (\xi, \xi')$ is written as
  \be  {\cal K}_t (\xi, \xi') 
		=\sqrt{\rho(\xi) \, \rho(\xi')} \,\frac{(1 - \gamma)^\beta 
		(1 - t)^{\xi+\xi'}}{(1 - \gamma t)^{\xi+\xi'+\beta}}\,    
		{}_2F_1 \left [ -\xi, -\xi'; \beta
			; \frac{t (1 - \gamma^2)}{\gamma (1 - t)^2} \right ] \,.
						\label{4.6}  \ee
For integer $\xi$ and $\xi'$ we have the limit 
  \be  \lim_{t \to 1^-} {\cal K}_t (\xi, \xi') 
			= \delta_{\xi,\xi'} \,. \label{4.7}  \ee
In this limit, the relation (\ref{4.2}) coincides with the 
dual orthogonality (\ref{2.9}) of the Meixner functions.

	 The limit of (\ref{4.6}) when $t \to i$ ($\tau\to\frac12\pi$)
corresponds to a 
discrete analogue of the classical Fourier-Bessel transform; whereas
the latter integrates over the nonnegative half-axis, the former sums
over the integer points $\xi=0,1,\ldots$. The limit is
  \bea 
	{\cal K}_i (\xi, \xi')&=& \displaystyle  
			\lim_{t \to i} {\cal K}_t (\xi, \xi') 
						\nonumber \\
		&=& \displaystyle \sqrt{\rho(\xi)\, \rho(\xi') }\,
			\frac{(-2i)^{(\xi+\xi')/{2}} 
			(1 - \gamma)^\beta}{(1 - i \gamma)^{\xi+\xi'+\beta}} \,
			{}_2F_1 \left [ -\xi, -\xi'; -\beta; 
			-\frac{(1-\gamma)^2}{2\gamma} \right ]\,. \label{4.8}
   \eea
It is easy to verify that for integer $\xi$ and $\xi''$,
  \be \ssp {\cal K}_i (\xi, \xi') \, {\cal K}_{i} (\xi', \xi'') 
					= \delta_{\xi,\xi''}\,. \label{4.9}  \ee


\section{Limit and special cases}
The difference model of the Meixner linear harmonic oscillator family
(\ref{2.7}) contains as limit and particular cases all the models of
Hermite (\ref{1.1}), Charlier (\ref{1.7}) and Kravchuk (\ref{1.18}).
We make these limits explicit below. Here we discuss also the
corresponding relations with the radial part of the nonrelativistic
Coulomb system in quantum mechanics and a relativistic model of the
linear oscillator, built in terms of the continuous Meixner-Pollaczek
polynomials. 

\subsection{Meixner $\longrightarrow$ Hermite}
From the recurrence relation for the Meixner polynomials (\ref{2.4}), it
can be show that the following limit to the Hermite polynomials holds:
  \be \lim_{\nu \to \infty} (2 \nu)^{n/2} \, 
		M_n \left( \frac{\nu + \sqrt{2 \nu} \xi}{1 - \gamma};
				\frac{\nu}{\gamma} , \gamma \right) 
		= (-1)^n \, H_n (\xi)\,. \label{5.1}  \ee
Furthermore, measures and normalization coefficients relate as 
  \bea
  \lim_{\nu \to \infty} \sqrt{2 \nu} \, (1 - \gamma)^{\nu/\gamma - 1}
	\rho \left( \frac{\nu + \sqrt{2 \nu} \xi}{1 - \gamma} \right) 
		&=& \displaystyle  \frac{1}{\sqrt{\pi}} \, e^{- \xi^2} \,,  \label{5.2}  \\
   \lim_{\nu \to \infty} (2 \nu)^{n/2} \,(1 - \gamma)^{\nu/2\gamma}
   d_n		&=& \displaystyle   \sqrt{2^n n!}\,. \label{5.3}  \eea

    The wavefunctions (\ref{2.6}) with argument 
$(\nu + \sqrt{2 \nu} \xi)/(1 - \gamma) $ and $ \beta = \nu/\gamma$,
coincide in the limit $\nu \to \infty$ with the wavefunctions of
the linear harmonic oscillator (\ref{1.3}), {\it i.e.},
  \be  \lim_{\nu \to \infty} \frac{(2 \nu)^{1/4}}{(1 - \gamma)^{1/2}}\,
	\psi_n \bigg( \frac{\nu + \sqrt{2 \nu} \xi}{1 - \gamma }
			; \frac{\nu}{\gamma}, \gamma \bigg)
				= \psi_n^\ssr{H} (\xi)\,. \label{5.4}  \ee
The combination $K_0(\xi) - \nu/{2\gamma}$ reproduces, in the same
limit, the product $a^+(\xi) a(\xi)$, whereas the matrix elements of
$\sqrt{\gamma/\nu} K_\pm (\xi)$ converge to the creation and 
annihilation operators $a^+(\xi)$ and $a(\xi)$, respectively. The
Meixner oscillator family (\ref{2.7}) thus contains as a limit case
the linear harmonic oscillator (\ref{1.1}) of quantum mechanics.

\subsection{Meixner $\longrightarrow$ Charlier}
It is known that the Meixner (\ref{2.1}) and Charlier (\ref{1.6})
polynomials are connected by the limit relation \cite{2,3}
  \be \lim_{\beta \to \infty} \, M_n (\xi; \beta, \mu/\beta) 
		= C_n (\xi, \mu)\,. \label{4.10}  \ee
Hence in the limit when $\beta \to \infty$ and $ \gamma = \mu/\beta
\to 0$, from (\ref{4.8}) one obtains the reproducing kernel for the
Charlier functions \cite{12}:  
  \be {\cal K}^\ssr{C} (\xi, \xi') 
	= \lim_{\beta \to \infty,\ \beta \gamma = \mu} {\cal K}_i (\xi, \xi')
	= e^{-(1-i)\mu} \, \sqrt{\frac{(-2i\mu)^{\xi+\xi'}}{{\xi!\,\xi'!}}} \, 
		{}_2F_0 \left[-\xi, -\xi'; -\frac{1}{2\mu}\right]. \label{4.11}  \ee

Using the limit relation (\ref{4.10}) it is easy to check that
  \be \lim_{\beta \to \infty} \, 
		\psi_n (\mu + \xi/h_1 ; \beta, \mu/\beta ) 
		=\psi_n^\ssr{C} (\xi) \,. \label{5.5}  \ee
Hence the wavefunctions (\ref{2.6}) with argument $\mu + \xi/h_1$ and
parameter $\gamma = \mu/\beta$ coincide, in the limit when $\beta \to
\infty$, with the wave funcitons of the difference (discrete) model
of the Charlier oscillator (\ref{1.7}). In the same limit, the
combination $H(s) + \onehalf(1 - \beta)$, where $s = \mu + \xi/h_1$, 
reproduces $H^\ssr{C} (\xi)/\hbar \omega$, whereas $\beta^{-1/2} K_\pm (\xi)$ 
tend to the raising and lowering operators $B^+$ and $B$, respectively.

\subsection{Meixner $\longrightarrow$ Kravchuk}
The Kravchuk polynomials (\ref{1.17}) are also a particular case of
the Meixner  polynomials (\ref{2.1}), with the parameters $\beta = -
N$ and $\gamma = - p/(1-p)$, that is, 
  \be K_n (\xi; p, N) = M_n (\xi; -N, p/(p - 1)) \,. \label{4.12}  \ee
In this case, from (\ref{4.8}) one obtains the reproducing 
kernel for the Kravchuk functions (\ref{1.19}) \cite{12},
  \be  {\cal K}^\ssr{K} (\xi, \xi') 
	= \sqrt{(-2 i p q)^{\xi+\xi'}\, C_N^\xi C_N^{\xi'}} \, 
			\Big(1 -(1-i)p\Big) ^{N-\xi-\xi'} \,   
			{}_2F_1 \left[-\xi, -\xi'; -n; \frac{1}{2p(1-p)}\right] 
								\label{4.13}  \ee 

As follows from the relation (\ref{4.12}) for $\beta = - N$,  $\gamma
= - p/(1-p)$ and argument $p N + \sqrt{2 p(1-p)N} \xi$, the model
(\ref{2.7}) coincides with the difference model of the Kravchuk 
oscillator (\ref{1.18}).

\subsection{Meixner $\longrightarrow$ Laguerre}
	The limit relation \cite{3,17}
\be
	\lim_{h\to0} M_n(x/h;\beta,1-h)=\frac{n!}{(\beta)_n}
			L_n^{\beta-1}(x), \label{5.6}
\ee
where $L_n^{\alpha}(x)$ are the Laguerre polynomials, enables us to
consider the nonrelativistic Coul\-omb system as another limit case of
the difference model (\ref{2.7}). Indeed, from  (\ref{2.6}) and
(\ref{5.6}), it follows that
\be
	\lim_{h\to0} \frac{1}{\sqrt{h}}\psi_n(2kr/h;2l+2,1-h)
		=\sqrt{2kr} R_{n,l}(2kr), \label{5.7}
\ee
where
\be
	R_{n,l}(x)=(-1)^n \sqrt{\frac{n!}{(n+2l+1)!}}
			x^l e^{-x/2} L_n^{2l+1}(x) \label{5.8}
\ee
is the radial wavefunction of the Coulomb system (see, for example,
\cite{18}), $r$ is the radial variable, $l$ and $n+l+1$ are orbital
and principal quantum numbers respectively, and
$k=me^2/\hbar^2(n+l+1)$. 
	
	In the same limit $h \to0$, the generators $K_0(x)$ and
$K_\pm(x)$, where $x = 2kr/h$ reproduce the well-known generators 
of the dynamical symmetry algebra su(1,1) for the nonrelativistic 
Coulomb model \cite{5},
\bea
	J_0 &=& \displaystyle -\frac{1}{2k}
		\left[ r\partial_r^2 + \partial_r -
			\frac{(l+\onehalf)^2}{r} - k^2r\right], \label{5.9a} \\
    J_\pm &=& -J_0 + kr \mp (r\partial_r +\onehalf). \label{5.9b}
\eea

	This connection between the Meixner polynomials and radial
wavefunctions for the nonrelativistic Coulomb system can be used for
constructing a $q$-analogue of the Coulomb wavefunctions in terms of
the $q$-Meixner polynomials (see Refs.\ \cite{19,20}).

\subsection{Meixner-Pollaczek (relativistic) oscillators}
	 There is the family of Meixner-Pollaczek polynomials
\be
	P_n^\lambda(x;\phi) = \frac{(2\lambda)_n}{n!}\,
			e^{-in\phi}\, F(-n,\lambda-ix;2\lambda;1-e^{2i\phi}),
							\label{Pollaczek}
\ee
which satisfy the orthogonality relation 
\be
	\int_{-\infty}^\infty 
			P_n^\lambda(\xi;\phi)\,
			P_{n'}^\lambda(\xi;\phi)\,\rho^\ssr{P}(\xi)\,d\xi\, 
			=\delta_{n,n'}\,\frac{\Gamma(2\lambda+n)}{n!}
			\label{Porthogonality} 
\ee
with respect to a continuous measure with the weight 
\be
	\rho^\ssr{P}(\xi)=\frac{1}{2\pi}(2\sin\phi)^{2\lambda}
		|\Gamma(\lambda+i\xi)|^2 \exp[(2\phi-\pi)\xi].
		\label{Pmeasure} 
\ee

	The reason why we mention these polynomials here is the following.
In Ref.\ \cite{27} it was shown that the Meixner polynomials 
$M_n(\xi;\beta,\gamma)$ and the 
Meixner-Pollaczek polynomials (\ref{Pollaczek}) are in fact
interrelated by
\be
	P_n^\lambda(\xi;\phi) = \frac{e^{-in\phi}}{n!} (2\lambda)_n
			M_n(i\xi-\lambda;2\lambda,e^{-2i\phi}).
			\label{Meixner-Pollaczek} 
\ee
The transition from the discrete orthogonality (\ref{2.2}) for the Meixner
polynomials $M_n(\xi;\beta,\gamma)$ to the continuous one
(\ref{Porthogonality}) is analogous to the well-known
Sommerfeld-Watson transformation in optics and quantum theory of
scattering. 

	In the relativistic model of the linear harmonic oscillator, proposed
in \cite{28}, the wavefunctions in configuration space
are expressed in terms of the Meixner-Pollaczek polynomials
(\ref{Pollaczek}) and their weight function (\ref{Pmeasure}) with the
specific value of the parameter $\phi=\onehalf\pi$. The same model in the
homogeneous external field $gx$ corresponds to the value of the 
parameter $\phi$ given by $\arccos(g/mc\omega)$, where $m$ and
$\omega$ have the same meaning as in the classical case, and $c$ is
the velocity of light \cite{28,29}. In other words, the relation\
(\ref{Meixner-Pollaczek}) gives the connection between 
the relativistic harmonic oscillator and the Meixner oscillator,
discussed in this paper.

\section*{Acknowledgements}
We thank the support of {\sc dgapa--unam} by the grant IN106595 {\it
Optica Matem\'atica}. One of us (Sh.M.N.) is grateful to {\sc
iimas--unam} for the hospitality extended to him during his visit to
Cuernavaca in April--June, 1997.

\def\jour#1#2#3#4{{\sl #1{}} {\bf #2}, #3\ #4}

\end{document}